\documentclass[twocolumn,prd,showpacs,preprintnumbers,amsmath,amssymb]{revtex4}

\usepackage{graphicx}
\usepackage{dcolumn}
\usepackage{bm}
\usepackage{epsfig}

\def\lan{\langle}
\def\ran{\rangle}

\def\aq{\bar{q}}
\def\beq{\begin{equation}}
\def\ee{\end{equation}}
\def\eeq{\end{equation}}

\def\vphi{\varphi}

\def\bfig{\begin{figure}}
\def\efig{\end{figure}}
\def\bea{\begin{eqnarray}}
\def\bwt{\begin{widetext}}
\def\ewt{\end{widetext}}
\def\beann{\begin{eqnarray*}}
\def\eea{\end{eqnarray}}
\def\eeann{\end{eqnarray*}}

\def\nn{\nonumber}

\def\raw{\rightarrow}

\def\3p0{$^{3}P_{0}$}

\begin{document}

\preprint{APS/123-QED}

\title{Meson decay in a corrected \3p0 model}

\author{D. T.  da Silva}
\affiliation{Instituto de F\'{\i}sica, Universidade Federal do 
Rio Grande do  Sul\\
 Av. Bento Gon\c{c}alves, 9500, Porto Alegre, 
Rio Grande do Sul, CEP 91501-970, Brazil}

\author{M. L. L. da Silva}
\affiliation{Instituto de F\'{\i}sica, Universidade Federal do Rio
  Grande do  Sul\\
 Av. Bento Gon\c{c}alves, 9500, Porto Alegre, Rio Grande do Sul,
CEP 91501-970, Brazil}

\author{J. N. de Quadros}
\affiliation{Instituto de F\'{\i}sica, Universidade Federal do Rio
  Grande do  Sul\\
 Av. Bento Gon\c{c}alves, 9500, Porto Alegre, Rio Grande do Sul,
CEP 91501-970, Brazil}

\author{D. Hadjimichef\,}
\email{dimiter@ufpel.edu.br; dimihadj@gmail.com}
\affiliation{Instituto de F\'{\i}sica e Matem\'atica, 
Universidade Federal de  Pelotas\\
 Campus Universit\'ario,  Pelotas, Rio Grande do Sul, CEP 96010-900, Brazil}

\date{\today}

\begin{abstract}

Extensively applied to both light and heavy  meson decay 
and standing as  one of the most successful strong decay models is the \3p0 model, in which 
$q\bar{q}$ pair production is the dominant mechanism. The pair production can be obtained 
from the  non-relativistic limit of a microscopic  interaction Hamiltonian involving Dirac quark fields.
The evaluation of the decay amplitude can be performed by a diagrammatic technique
for drawing quark lines.  In this paper we use an alternative approach
which consists  in a mapping technique, 
the Fock-Tani formalism, in order to obtain an effective Hamiltonian 
starting from same  microscopic  interaction.
An additional effect is manifest in this formalism associated to the 
extended nature of mesons: bound-state
corrections. A corrected \3p0 is obtained and applied, as an example, 
to  $b_{1}\rightarrow\omega\pi$ and $a_{1}\rightarrow\rho\pi$ decays.

\end{abstract}



\maketitle


\section{Introduction}
\label{intro}

A great variety of quark-based models are known  that  describe with
reasonable success single-hadron properties. A natural  question
that arises is to what extent a model which gives a good description
of hadron properties is, at the same time, able to describe the
complex hadron-hadron interaction or by the same principles hadron
decay. In particular, the theoretical aspects  of strong decay have
been challenged by QCD exotica (glueballs and hybrids) where a
consistent understanding  of the mixing schemes for these states is
still an open question \cite{close}-\cite{klempt2} . 
The nature of the family of ``new mesons''
$X,Y,Z$ \cite{barnes-xyz} is another unsolved puzzle: are they actually new $q\bar{q}$
mesons, hadronic molecules or something else?  In the  direction of
clarifying these questions
 is the successful
decay model, the \3p0 model, which considers only
OZI-allowed strong-interaction decays.
This model was introduced over thirty years ago by Micu
\cite{micu} and  applied to meson decays in the 1970 by LeYaouanc {\it et al.}
\cite{leyaouanc}. This description is a natural consequence of the
constituent quark model  scenario  of hadronic states. 

T. Barnes {\it  et al.} \cite{barnes1}-\cite{barnes4} have made an
extensive survey of meson states in the
light of the \3p0 model. Two basic parameters of their formulation are 
$\gamma$ (the interaction strength)  and $\beta$ (the wave function's
extension parameter). Although they found the optimum values near
$\gamma= 0.5$ and $\beta=0.4$ GeV, for light 1S and 1P  decays, these 
values lead to overestimates of the widths of\,\, higher-L states. In
this perspective a modified $q\bar{q}$ pair-creation interaction, with
 $\gamma= 0.4$ was preferred.

In the present work, we employ a mapping technique in order
to obtain an effective interaction for meson decay.
A particular mapping technique
long used in atomic physics \cite{girar1}, the Fock-Tani formalism
(FTf), has been adapted, in previous publications \cite{annals}-\cite{mario}, in order
to describe hadron-hadron scattering interactions with constituent
interchange. Now this technique has been extended in order to
include meson decay.
We start from the microscopic $q\bar{q}$ pair-creation interaction, as will be shown,
in lower order, the \3p0 results are reproduced. An additional and
interesting feature appears in higher orders of the formalism: corrections due to
the bound-state nature of the mesons and a natural modification
in the $q\bar{q}$ interaction strength.

In the Fock-Tani formalism  one starts with the Fock representation of the
system using field operators of elementary constituents which satisfy canonical
(anti) commu\-ta\-tion relations. Composite-particle field operators  are
linear combinations of the elementary-particle operators and do not generally
satisfy canonical (anti) commutation relations. ``Ideal" field
operators acting on an enlarged Fock space are then introduced in close
correspondence with the composite ones.
Next, a given unitary transformation, which transforms the
single composite states into single ideal states, is introduced.
Application of the
unitary operator on the microscopic Hamiltonian, or on
other hermitian operators expressed in terms of the elementary constituent
field operators, gives equivalent operators
which contain the ideal field operators. The effective Hamiltonian
in the new representation
has a clear physical
interpretation in terms of the processes it describes. Since all field
operators in the new representation satisfy canonical (anti)commutation
relations, the standard methods of quantum field theory can then be readily
applied.

In this paper we shall extend the FTf  to meson decay processes.
In the next section we review the basic aspects of the formalism.
Section \ref{sec:3p0-ftf} is dedicated to obtain an effective
decay Hamiltonian. 
In section \ref{sec:examples}, two light mesons decays  examples
are calculated  $b_{1}\rightarrow\omega\pi$ and
$a_{1}\rightarrow\rho\pi$. The summary and conclusions are followed
by appendixes which detail the method employed throughout this work.


\section{Mapping of mesons}
\label{sec:mesons}

This section reviews the formal aspects of the mapping procedure and
how it is implemented  to quark-antiquark meson states \cite{annals}.
The starting point of the Fock-Tani formalism  is the definition of single
composite bound states. We write a single-meson state in terms of a meson
creation operator $M_{\alpha}^{\dagger}$ as
\bea
|\alpha \ran  = M_{\alpha}^{\dagger}|0 \ran ,
\label{1b}
\eea
where $|0 \ran$ is the vacuum state. The meson creation operator $M_{\alpha}^{\dagger}$ is written
in terms of constituent quark and antiquark creation operators
$q^{\dagger}$ and $\aq^{\dagger}$,
\bea
M^{\dagger}_{\alpha}= \Phi_{\alpha}^{\mu \nu}
q_{\mu}^{\dagger} {\aq}_{\nu}^{\dagger} ,
\label{Mop}
\eea
$\Phi_{\alpha}^{\mu \nu}$ is the meson wave function
and $q_{\mu}|0 \ran=\aq_{\nu}|0\ran=0$.
The index $\alpha$ identifies the meson quantum numbers 
of space, spin and isospin. The indices $\mu$ and $\nu$ denote
the spatial, spin, flavor, and color quantum numbers of the constituent
quarks. A sum over repeated indices is implied. It is convenient to work
with orthonormalized amplitudes,
\bea
\lan\alpha|\beta\ran= \Phi_{\alpha}^{*\mu \nu}
\Phi_{\beta}^{\mu \nu}=\delta_{\alpha \beta}.
\label{norm}
\eea
The quark and antiquark operators satisfy canonical anticommutation relations,
\bea
&&\{q_{\mu}, q^{\dagger}_{\nu}\}=
\{\aq_{\mu},\aq^{\dagger}_{\nu}\}=\delta_{\mu \nu}, \nn\\
&&\{q_{\mu}, q_{\nu}\}=\{\aq_{\mu},\aq_{\nu}\}=
\{q_{\mu}, \aq_{\nu}\}= \{q_{\mu}, \aq_{\nu}^{\dagger}\}=0.
\label{qcom}
\eea
Using these quark anticommutation relations, and the normalization condition
of Eq.~(\ref{norm}), it is easily shown that the meson operators satisfy the
following non-canonical commutation relations
\beq
[M_{\alpha}, M^{\dagger}_{\beta}]=\delta_{\alpha \beta} -
\Delta_{\alpha \beta},\hspace{1.5cm}[M_{\alpha}, M_{\beta}]=0,
\label{Mcom}
\eeq
where
\bea
\Delta_{\alpha \beta}= \Phi_{\alpha}^{*{\mu \nu }}
\Phi_{\beta}^{\mu \sigma }\aq^{\dagger}_{\sigma}\aq_{\nu}
+ \Phi_{\alpha}^{*{\mu \nu }}
\Phi_{\beta}^{\rho \nu}q^{\dagger}_{\rho}q_{\mu}.
\label{delta}
\eea
In addition,
\bea
&&[q_{\mu},M_{\alpha}^{\dagger}]=\Phi^{\mu \nu}_{\alpha}
{\aq}_{\nu}^{\dagger}\hspace{.5cm},\hspace{.5cm}
[{\aq}_{\nu},M_{\alpha}^{\dagger}]=-\Phi^{\mu \nu}_{\alpha}
q_{\mu}^{\dagger},\hspace{1.0cm}\nn\\
&&[q_{\mu},M_{\alpha}]=[{\aq}_{\nu}, M_{\alpha}]=0.
\label{MMq}
\eea

The presence of the operator $\Delta_{\alpha \beta}$ in Eq.~(\ref{Mcom}) is
due to the composite nature of the mesons. This term enormously complicates
the mathematical description of processes that involve the
hadron and quark degrees of freedom. The usual field theoretic techniques
used in many-body problems, such as the Green's functions method, Wick's
theorem, etc, apply to creation
and annihilation operators that satisfy canonical relations. Similarly, the
non-vanishing of the commutators $[q_{\mu},M_{\alpha}^{\dagger}]$ and
$[{\aq}_{\nu},M_{\alpha}^{\dagger}]$ is a manifestation of the lack of
kinematic independence of the meson operator from the quark and antiquark
operators. Therefore, the meson operators $M_{\alpha}$ and
$M_{\alpha}^{\dagger}$ are not convenient dynamical variables to be used.

A transformation is defined such that  a single-meson state
$|\alpha \ran$ is redescribed by an (``ideal") elementary-meson state by
\bea
|\alpha \ran\longrightarrow U^{-1}|\alpha\rangle =
m^{\dagger}_{\alpha}|0\rangle,
\label{single_mes}
\eea
where $m^{\dagger}_{\alpha}$ an ideal meson creation operator. The ideal
meson operators $m^{\dagger}_{\alpha}$ and $m_{\alpha}$ satisfy,
by definition, canonical commutation relations
\beq
[m_{\alpha}, m^{\dagger}_{\beta}]=\delta_{\alpha \beta} ,
\hspace{1.5cm}[m_{\alpha}, m_{\beta}]=0.
\label{mcom}
\eeq
The state $|0\rangle$ is the vacuum of both $q$ and $m$ degrees of freedom in the
new representation.
In addition, in the new representation the quark and antiquark operators
$q^{\dagger}$, $q$, $\aq^{\dagger}$ and $\aq$ are kinematically independent of
the $m^{\dagger}_{\alpha}$ and $m_{\alpha}$
\bea
[q_{\mu},m_{\alpha}]=[q_{\mu},m^{\dagger}_{\alpha}]=[\aq_{\mu},m_{\alpha}]=
[\aq_{\mu},m^{\dagger}_{\alpha}]=0 . 
\label{indep_mes} 
\eea 
The unitary operator $U$ of the transformation is 
\bea
U(t)=\exp\left[ t\, F\right] ,
\label{u} 
\eea 
where $F$ is the generator of the transformation and $t$ a parameter which is set to
$-\pi/2$ to implement the mapping. The next step is to obtain the
transformed operators in the new representation. The basic operators
of the model are expressed in terms of the quark operators. Therefore, in
order to obtain the  operators  in the new representation,
one writes
\bea
q(t)=U^{-1}\, q \,U,\hspace{1.0cm}{\aq}(t)=U^{-1}\, {\aq}\, U .
\eea
The generator $F$ of the transformation is
\bea
F= m^{\dag}_{\alpha}\,\tilde{M}_{\alpha}-
\tilde{M}^{\dag}_{\alpha} m_{\alpha}
\label{f-generator}
\eea
where
\bea
\tilde{M}_{\alpha}=\sum_{i=0}^{n}\tilde{M}^{(i)}_{\alpha},
\label{mes_gen}
\eea
with
\bea
&&[\tilde{M}_{\alpha},\tilde{M}^{\dagger}_{\beta}] = \delta_{\alpha\beta}
\hspace{.5cm} + \hspace{.5cm}{\cal O} (\Phi^{n+1}),\nn\\
&&[\tilde{M}_{\alpha}, \tilde{M}_{\beta}]=
[ \tilde{M}^{\dagger}_{\alpha},  \tilde{M}^{\dagger}_{\beta}]=0.
\label{comO}
\eea
It is easy to see from (\ref{f-generator}) that $F^{\dag}=-F$ which
ensures that $U$ is unitary.
The index $i$ in (\ref{mes_gen}) represents the order of the expansion in powers of
the wave function $\Phi$. The $\tilde{M}_{\alpha}$ operator is determined up
to a specific order $n$ consistent with (\ref{comO}). The examples
studied in \cite{annals} required the determination of
$\tilde{M}^{(i)}_{\alpha}$  up to order 3 as shown below
\bea
\tilde{M}_{\alpha}^{(0)}&=&M_{\alpha}  \hspace{.5cm};  \hspace{.5cm}
\tilde{M}_{\alpha}^{(1)}= 0  \nn\\
\tilde{M}_{\alpha}^{(2)} &=&  \frac{1}{2}\Delta _{\alpha  \beta}M_{\beta}
  \hspace{.1cm};  \hspace{.1cm}
\tilde{M}_{\alpha}^{(3)} =\frac{1}{2}M^{\dagger}_{\beta}\,\,
T_{\alpha\beta\gamma} \,\,M_{\gamma}, 
\label{mes_gen2} 
\eea with$T_{\alpha\beta\gamma} =-\left[M_{\alpha},\,\Delta_{\beta\gamma}\right]$.  In the
 ``zero-order" approximation,  overlap among  mesons
is neglected and terms of the same power in the bound-state wave
function $\Phi_{\alpha}$ ($\Phi^{*}_{\alpha}$) are collected. 
In order to have a consistent power counting scheme, the
implicit $\Phi_{\alpha}$ ($\Phi^{*}_{\alpha}$) entering via
Eq.~(\ref{Mop}) are not counted.  The consequence of this is that
the equations for $m_{\alpha}$ and $\tilde{M}_{\alpha}$ are
manifestly symmetric,
\bea
{d m_{\alpha}(t)\over dt}&=&[m_{\alpha}(t),F]
= \tilde{M}_{\alpha}(t),\nn\\
{d \tilde{M}_{\alpha}(t)\over dt}&=&[\tilde{M}_ {\alpha}(t),F]=-m_{\alpha}(t),
\label{solm&O}
\eea
and their solutions involve only trigonometric functions of $t$,
\bea
m_{\alpha}(t)&=& \tilde{M}_{\alpha} \sin t + m_{\alpha} \cos t , \nn\\
\tilde{M}_{\alpha}(t)&=& \tilde{M}_{\alpha} \cos t - m_{\alpha} \sin t.
\label{solOm}
\eea
The equations of motion for the quark operators $q$ and $\aq$ can be obtained
by making use of Eq.~(\ref{MMq}) in a similar way,
\bea 
{dq_{\mu}(t) \over dt }&=& \left[q_{\mu}(t), F\right]
\hspace{.5cm};\hspace{.5cm} 
{d\aq_{\mu}(t) \over dt } =
\left[\aq_{\mu}(t), F\right]. 
\label{eqq} 
\eea 
In the zero-order approximation, the effects of the meson structure are
neglected resulting
\bea
&&q^{(0)}_{\mu}(t)=q_{\mu},
\hspace{0.5cm}
\aq^{(0)}_{\nu}(t)=\aq_{\nu}, \nn\\
&&m^{(0)}_{\alpha}(t)=m_{\alpha}\cos t + M_{\alpha}\sin t,
\nn\\
&&M^{(0)}_{\alpha}(t)=M_{\alpha}\cos t - m_{\alpha}\sin t.
\label{sol0}
\eea
In first order one has
\bea
&&q^{(1)}_{\mu}(t) = -\,\Phi^{\mu \nu_{1}}_{\alpha}
\aq^{\dagger }_{\nu_{1}} \left[m_{\alpha} \sin t + M_{\alpha}
\left(1 - \cos t \right)\right],\nn\\
&&\aq^{(1)}_{\nu}(t) = \Phi^{\mu_{1}\nu }_{\alpha}
q^{\dagger }_{\mu_{1}} \left[ m_{\alpha} \sin t + M_{\alpha} \left(1 - \cos t\right)
\right],\nn\\
&&m^{(1)}_{\alpha}(t)=0,\hspace{1.0cm}M^{(1)}_{\alpha}(t)=0.
\label{q1}
\eea
The second and  third order solutions to (\ref{eqq})  were calculated
in reference \cite{annals} and appear again, for completeness, in
appendix \ref{operators}, together with the higher order operators  required in our
calculation.

Once a microscopic interaction Hamiltonian $H$ is defined, at the quark
level, a new transformed Hamiltonian can be obtained. This effective
interaction we shall call the {\sl Fock-Tani Hamiltonian} and is
evaluated  by the application of the unitary operator
$U$ on the microscopic Hamiltonian ${\cal H}_{\rm FT}=U^{-1}HU$.
The transformed Hamiltonian ${\cal H}_{\rm FT}$ describes all possible
processes involving mesons and quarks.
The general structure of ${\cal H}_{\rm FT}$  is of  the following form
\bea
{\cal H}_{\rm FT}={\cal H}_{\rm\bf q} + {\cal H}_{\rm\bf m} + {\cal H}_{\rm\bf m q} ,
\label{separation}
\eea
where the first term involves only quark operators, the second one involves
only ideal meson operators, and ${\cal H}_{\rm\bf m q}$ involves quark and
meson operators.

In   ${\cal H}_{\rm FT}$ there are higher order terms
that provide bound-state corrections
(also called orthogonality corrections) to the lower order ones. The
basic quantity for these corrections is the {\it bound-state kernel}
  $\Delta(\rho\tau;\lambda\nu)$ defined as
\begin{eqnarray}
\Delta(\rho\tau;\lambda\nu)
=\Phi^{\rho\tau}_{\alpha}\Phi^{\ast\lambda\nu}_{\alpha}.
\label{kernel}
\end{eqnarray}

To discuss the physical meaning of the
bound-state corrections and how they modify the 
fundamental quark interaction we shall present
an example, in a toy model similar to the model studied in \cite{annals}, where the basic arguments are outlined.
In this example, the  starting point is a
two-body microscopic quark-antiquark Hamiltonian of the form 
\bea
H_{2q} &=& T\left(\mu\right)q^{\dagger}_{\mu} q_{\mu} + T\left(\nu\right)
\aq_{\nu}^{\dagger}\aq_{\nu} 
+ V_{q\aq}(\mu\nu;\sigma\rho)q^{\dagger}_{\mu}\aq^{\dagger}_{\nu}
\aq_{\rho}q_{\sigma} 
\nn\\
&&
\!\!\!\!
+ \frac{1}{2} V_{qq}(\mu\nu;\sigma\rho)
q^{\dagger}_{\mu}q^{\dagger}_{\nu}q_{\rho}q_{\sigma}
+\frac{1}{2}V_{\aq\aq}(\mu\nu;\sigma\rho)\aq^{\dagger}_{\mu}
\aq^{\dagger}_{\nu}\aq_{\rho}\aq_{\sigma}.
\nn\\
\label{qHamilt}
\eea
The transformation ${\cal H}_{\rm FT}=U^{-1}\,H_{2q}\, U$ is implemented again by  
transforming each quark and antiquark 
operator in Eq.~(\ref{qHamilt}), where a  similar structure to
Eq. (\ref{separation}) is obtained.
In free space, the wave function $\Phi$ of Eq.~(\ref{Mop}) satisfy the following 
equation
\beq
H(\mu\nu; \sigma\rho)\Phi_{\alpha}^{\sigma\rho}=\epsilon_{[\alpha]} 
\Phi_{[\alpha]}^{\mu \nu},
\label{Schro}
\eeq
where $H(\mu\nu; \sigma\rho)$ is the Hamiltonian matrix
\bea
H(\mu\nu; \sigma\rho)&=&\delta_{\mu[ \sigma]} \delta_{\nu[\rho]} 
\left[T([\sigma]) + T([\rho])\right]
\nn\\
&&+ 
V_{q\aq}(\mu\nu; \sigma\rho) ,
\label{Schroperat}
\eea
$\epsilon_{[\alpha]}$ is the total energy of the meson. There is no sum 
over repeated indices inside square brackets.

The  effective quark Hamiltonian ${\cal H}_{\rm\bf q}$  has an identical
structure to the  microscopic quark 
Hamiltonian, Eq.~(\ref{qHamilt}), except that the term corresponding to the 
quark-antiquark interaction is modified as follows
\bea
\!\!\!\!
{\cal V}_{q\bar{q}} &=&\left[\,V_{q\aq}-H\,\Delta - \Delta\,H    
 +  \Delta\,H\,\Delta\,\right]\, ,
\label{modqaq}
\eea
where $V_{q\bar{q}} \equiv V_{q\aq}(\mu \nu; \sigma \rho)$
and the contraction 
$H\,\Delta\equiv H(\mu \nu;\tau\xi)\,\Delta(\tau\xi; \sigma \rho)$.
An important  property of the bound-state kernel is 
\beq
\Delta(\mu \nu; \sigma\rho)\Phi^{\sigma\rho}_{\alpha}=\Phi^{\mu\nu}_{\alpha},
\label{propDelta}
\eeq
which follows from the wave function's orthonormalization, Eq.~(\ref{norm}).
In the case that $\Phi$ is a solution of
Eq.~(\ref{Schro}), the new quark-antiquark interaction term becomes 
\bea
{\cal V}_{q\aq}(\mu \nu ; \sigma \rho) =
V_{q\aq}(\mu \nu ; \sigma \rho) - \sum_{\alpha}
\epsilon_{\alpha}\Phi^{\ast \mu \nu}_{\alpha} \Phi^{\sigma \rho}_{\alpha}.
\label{modV}
\eea
The spectrum of the modified quark 
Hamiltonian, ${\cal H}_{\rm\bf q}$, is positive semi-definite and hence has
no bound-states~\cite{girar1}. 
This result is exactly the same as in Weinberg's 
quasiparticle method~\cite{quasi}, where the bound-states 
are redescribed by  ideal particles.  The new ${\cal V}_{q\aq}$ is a weaker  potential, modified in such a way
that no new bound-states are formed.

In the quark-meson sector of Eq. (\ref{separation}) in  ${\cal H}_{\rm\bf m q}$  appears a term
related to spontaneous meson break-up
\bea
H_{\rm m\to q \bar{q}}=V(\mu\nu;\alpha)\, 
q^{\dagger}_{\mu}\aq^{\dagger}_{\nu}m_{\alpha}
\label{qqm}
\eea
with
\bea
V(\mu\nu;\alpha)&=&H(\mu\nu;\sigma\rho)\Phi^{\sigma\rho}_{\alpha}
-\Delta(\mu\nu;\sigma\rho)H(\sigma\rho;\tau\lambda)\Phi^{\tau\lambda}_{\alpha}.
\nn\\
\label{v-qqm}
\eea
Again, in the case that $\Phi$ is a solution of Eq.~(\ref{Schro}),  a straightforward
calculation  demonstrates that $ H_{\rm m\to q \bar{q}} =0 $.  When there is no external interaction,  this result is a direct consequence
of the bound-state's stability against spontaneous break-up. This term can be interesting  in studies related to dense hadronic mediums. For
these systems  the wave function  is, in general,  not a solution of Eq.~(\ref{Schro}) and the strength of the potential $V(\mu\nu;\alpha)$ 
is now only decreased \cite{sergio}.

In the ideal meson sector ${\cal H}_{\rm \bf m}$ many similar approaches to FTf \cite{annals}  
have obtained the meson-meson scattering interaction in the Born
approximation: Resonating Group Method (RGM) \cite{oka}, Quark Born Diagram Formalism (QBDF) \cite{qbd},

\beq
H_{mm}= T_{mm} +V_{mm},
\eeq
where $T_{mm}$ is  the kinetic term and $V_{mm}$  is the meson-meson interaction potential with constituent interchange.
This potential is given by
\bea
V_{mm}=V_{mm}^{dir}+V_{mm}^{exc}+V_{mm}^{int} \,,
\eea
where $V_{mm}^{dir}$ is the direct 
potential (no quark interchange), $V_{mm}^{exc}$ the quark exchange term and $V_{mm}^{int}$ 
the intra-exchange term.  As shown in Ref. \cite{annals} and \cite{sergio}, if one extends the FT calculation to higher orders
a new meson-meson Hamiltonian is obtained
\bea
\bar{H}_{mm}=H_{mm}+\delta H_{mm}
\label{h-mm}
\eea
where $\delta H_{mm} $ is the bound-state correction Hamiltonian.
If the wave function  $\Phi$ is chosen to be an eigenstate of the microscopic quark 
Hamiltonian, then the intra-exchange term $V_{mm}^{int}$ is cancelled
\bea
V_{mm}^{int}+\delta H_{mm}=0.
\label{intra-cancel}
\eea

  In summary, these  examples reveal an important and common feature   of bound-state
corrections: they weaken the quark-antiquark potential.  In the next section we shall
follow the same procedure for a  quark pair creation interaction, which is fundamental
for the description of meson decay. Similar to the toy model, the  resulting interaction 
that describes meson decay, will contain a Born order contribution and  a bound-state correction.


\section{The \3p0 Decay Model in the Fock-Tani Formalism}
\label{sec:3p0-ftf}

In the paper of
 E. S. Ackleh, T. Barnes and E. S. Swanson \cite {barnes1}
a formulation of the \3p0 model is presented.
It  regards the  decay of an initial
state meson in the presence  of a $q\aq$ pair created from the vacuum.
The pair production is obtained from the  non-relativistic limit of
the  interaction Hamiltonian $H_{I}$ involving Dirac quark fields
\bea H_{I}=  2\,m_q\,\gamma\,\int d\vec{x}\,
\bar{\psi}(\vec{x})\,\psi(\vec{x})\,,
\label{3p0}
\eea
where $\gamma$ is the pair production strength.  For a $q\aq$ meson
$A$ to decay to mesons $B+C$ we must have $(q\aq)_A\to (q\aq)_B
+(q\aq)_C$. To determine the decay rate a matrix element of
(\ref{3p0}) is evaluated
\bea 
\langle B C | H_I | A \rangle =
\delta(\vec{P}_A-\vec{P}_B-\vec{P}_C)\, h_{fi}. 
\label{matrix} 
\eea
The evaluation of $h_{fi}$ is performed by diagrammatic technique
for drawing quark lines. The $h_{fi}$ decay amplitude is combined with
relativistic phase space, resulting in the differential decay rate
\bea
\frac{d\Gamma_{A\to BC}}{d\Omega}=2\pi\,P\,
\frac{E_B\,E_C}{M_A}|h_{fi}|^2
\label{dif-gamma}
\eea
which after integration in the solid angle $\Omega$ a  usual choice for the meson
momenta is made: $\vec{P}_A=0$ 
($P=|\vec{P}_B|=|\vec{P}_C|$).

In our approach,
the starting point for the Fock-Tani $h_{fi}$ is also the microscopic
Hamiltonian $H_I$ in (\ref{3p0}). The momentum expansion of the quark fields,
color and flavor are not represented explicitly, is
\begin{eqnarray}
  \psi(\vec{x}) = \sum_s \int \frac{d^3k}{(2\pi)^{3/2} }
[u_s(\vec{k})q_s(\vec{k}) \nn\\
 + v_s(-\vec{k})
\bar{q}_{s}{^\dag}(-\vec{k})]
  e^{i\vec{k}\cdot\vec{x}}\,.
\end{eqnarray}
In the product $\bar{\psi}(x)\psi(x)$ we shall
retain only the $q^{\dag}\bar{q}^{\dag}$ term, which yields from
Eq. (\ref{3p0}) a Hamiltonian in  a compact form,
\bea
H_{I}=V_{\mu\nu}\, q^{\dag}_{\mu}\aq^{\dag}_{\nu}
\label{h_3p0}
\eea
where sum (integration) is again  implied  over repeated indexes.
In the compact notation, the quark and antiquark momentum, spin,
flavor and color
 are written as
$\mu=(\vec{p}_{\mu},s_{\mu},f_{\mu},c_{\mu})$; 
$\nu=(\vec{p}_{\nu},s_{\nu},f_{\nu},c_{\nu})$, while the pair creation potential
$V_{\mu\nu}$ is given by
\bea 
V_{\mu\nu}\equiv 2\,m_{q}\, \gamma\,
\,\delta(\vec{p}_{\mu}+\vec{p}_{\nu})\, \bar{u}_{s_{\mu}f_{\mu}c_{\mu}  }
(\vec{p}_{\mu}) \, v_{s_{\nu} f_{\nu}c_{\nu}  }(\vec{p}_{\nu}) . 
\label{vmn} 
\eea
It should be noted that since Eq. (\ref{3p0})  is meant to be taken
in the nonrelativistic limit,  Eq. (\ref{vmn}) should be as well. In
the meson decay calculations, of the next section, this 
limit is considered.

\begin{figure*}[t]
\begin{center}
\epsfig{file=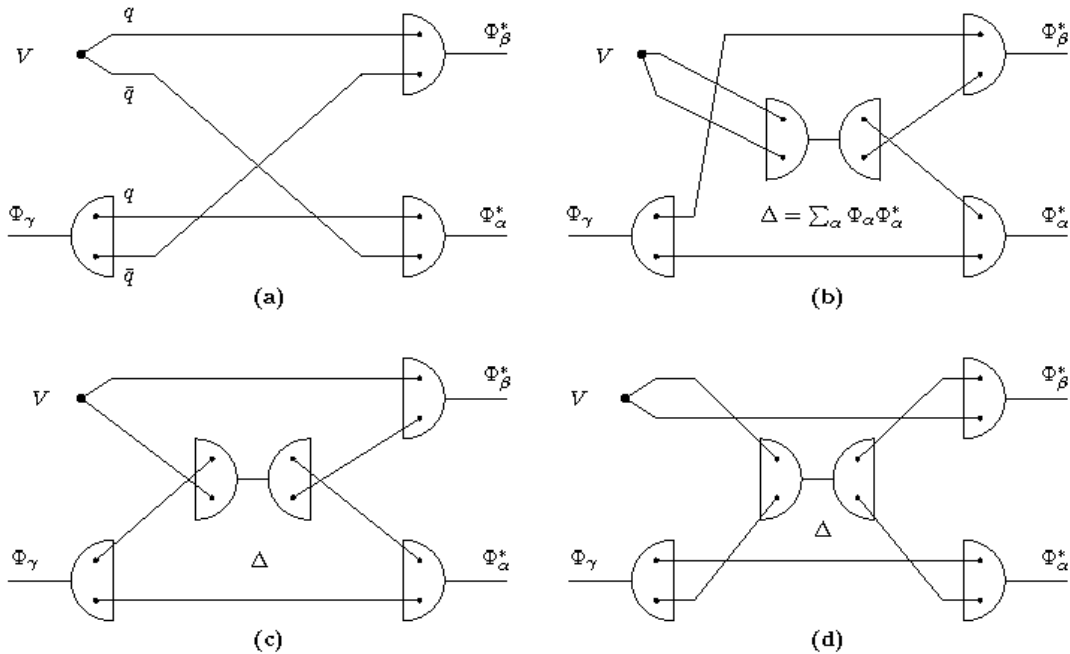, height=270.00pt, width=430.00pt}
\end{center}
\caption{Diagrams representing the C\3p0 model. Diagram (a)
  corresponds to the \3p0 amplitude. Diagrams (b), (c) and (d) are the
bound state corrections. The complete $h_{fi}$ amplitude includes
 the diagrams above plus diagrams with $\alpha\leftrightarrow \beta$.  }
\label{diag}
\end{figure*}

Applying the Fock-Tani transformation to $H_{I}$ one obtains the
effective Hamiltonian 
\bea 
{\cal H}_{FT}=U^{-1}\,\, H_{I} \,\,U.
\label{h_ft3p0} 
\eea
 The physical quantities in the FTf appear in a
second quantization notation. The effective decay amplitude will be
a product of the ideal meson operators with the  following structure
in the ideal meson sector: $m^{\dag}m^{\dag}m$. To obtain this
product corresponds to expand, in powers of the wave function, up to
third order. A Hamiltonian that describes this decay process, which
we shall call $H_{m}$, can be extracted from the mapping
(\ref{h_ft3p0}) by the following products
 \bea
 H_{m}=V_{\mu\nu}\,q^{\dag(3)}_{\mu}\aq^{\dag
(0)}_{\nu}+ V_{\mu\nu}\,q^{\dag(1)}_{\mu}\aq^{\dag (2)}_{\nu}.
\label{h_ft3p0b}
\eea
After the substitution of Eqs. (\ref{sol0}), (\ref{q1}),
(\ref{II.68}) and (\ref{II.70}) into (\ref{h_ft3p0b}) results in the
effective meson decay Hamiltonian \bea
H_{m}=f^{\mu\nu}(\alpha,\beta,\gamma)\,
V_{\mu\nu}\, m^{\dag}_{\alpha} m^{\dag}_{\beta} m_{\gamma}
\label{h_ft}
\eea
where
\bea
f^{\mu\nu}(\alpha,\beta,\gamma)=
-\Phi^{\ast\mu\tau}_{\alpha}
\Phi^{\ast\rho\nu}_{\beta}
\Phi^{\rho\tau}_{\gamma}.
\label{f}
\eea
In the ideal meson space the new initial and final states involve only ideal
meson operators $|A\rangle=m^{\dag}_{\gamma}|0\rangle$ and 
$|BC\rangle=m^{\dag}_{\alpha}m^{\dag}_{\beta} |0\rangle $.
The \3p0 amplitude is obtained in the FTf by an expression 
 equivalent to Eq. (\ref{matrix}),
\bea 
\!\!\!\!\!\!\!\!\!\!\!\!\!\!\!!\!\!
\langle B C | H_I | A \rangle &=&
\langle 0|m_{\alpha}m_{\beta} 
\, H_{m} \, m^{\dag}_{\gamma}|0\rangle
\nn\\
&=&f^{\mu\nu}(\alpha,\beta,\gamma)V_{\mu\nu}
+f^{\mu\nu}(\beta,\alpha,\gamma)V_{\mu\nu}.
\label{ideal-matrix} 
\eea
The term $f^{\mu\nu}(\beta,\alpha,\gamma)$
of (\ref{ideal-matrix})  is shown in Fig. (\ref{diag}a), the term
$f^{\mu\nu}(\alpha,\beta,\gamma)$ 
corresponds to the same diagram  with $\alpha\leftrightarrow \beta$.

In the FTf perspective a new 
 aspect is introduced  to meson decay: bound-state corrections. 
The lowest order correction is one that involves only one bound-state
kernel $\Delta(\mu\nu;\sigma\rho)$. This implies that the Hamiltonian
representing this correction
must be of fifth order in the power expansion of the wave function.

We shall call this new Hamiltonian, with
 the same basic operatorial structure
$m^{\dag}_{\alpha} m^{\dag}_{\beta} m_{\gamma}$, of
$\delta H_{m}$.
The only combinations $q^{\dag\,(i)}\aq^{\dag\,(j)}$ that results in a 
fifth order Hamiltonian  are 
\bea
\delta H_{m} \!\!&=&
\!\!\!
\left[ q^{\dag (3)}_{\mu}\aq^{\dag (2)}_{\nu}+ q^{\dag
(1)}_{\mu}\aq^{\dag (4)}_{\nu}+ q^{\dag (5)}_{\mu}\aq^{\dag
(0)}_{\nu}\right]V_{\mu\nu}.
\label{dh}
\eea
Details of this calculation is found in the appendix \ref{dhamil}. The 
bound-state corrected \3p0 Hamiltonian, which shall be called the
{\it $C^{\,3}P_0$ Hamiltonian}, is
\bea
\!\!\!\!\!\!\!\!\!\!\!\!\!\!\!\!
H^{\rm C3P0}&=&H_m+\delta H_m
\nn\\
&=&
-\Phi^{\ast\rho\xi}_{\alpha}
\Phi^{\ast\lambda\tau}_{\beta}
\Phi^{\omega\sigma}_{\gamma}\,
V^{\rm C3P0}\, m^{\dag}_{\alpha} m^{\dag}_{\beta} m_{\gamma}
\label{c3p0}
\eea 
where $V^{\rm C3P0} $ is a condensed notation for 
\bea
 V^{\rm C3P0}&=&
\left[\delta_{\mu\lambda}
\delta_{\xi\nu}
\delta_{\omega\rho}
\delta_{\sigma\tau}
-
\frac{1}{2}
\delta_{\sigma\xi}\,\delta_{\lambda\omega}\,\,
\Delta(\rho\tau;\mu\nu)
\right.
\nn\\
&&
+
\frac{1}{4}
\delta_{\sigma\xi}\,\delta_{\mu\lambda}\,\,
\Delta(\rho\tau;\omega\nu)
\nn\\
&&
+
\left.
\frac{1}{4}
\delta_{\xi\nu}\,\delta_{\lambda\omega}\,\,
\Delta(\rho\tau;\mu\sigma)
\right]V_{\mu\nu}.
\label{vc3p0}
\eea


\section{Light Meson Decay Examples}
\label{sec:examples}

The light meson sector is an interesting test ground where
the effects of the bound-state correction can be compared
to the usual \3p0 model. In particular, as examples, two specific decay
processes will be studied:  
$b_{1}\rightarrow\omega\pi$ and $a_{1}\rightarrow\rho\pi$. The wave
function and details of the matrix elements are found in the appendix 
\ref{wave-matrix}.
%
The  general decay amplitude can be written as
\bea
h_{fi}^{\rm C3P0} & = & \frac{\gamma}{\pi^{1/4}\,\beta^{1/2}}
\,{\cal  M}_{fi}^{\rm C3P0}.
\label{hfi-decay}
\eea
 For the first decay process, $b_{1}\rightarrow\omega\pi$,
 results in a decay amplitude given by
\begin{eqnarray}
{\cal M}_{fi}^{\rm C3P0} & = & 
{\mathcal{C}}_{01}\, Y_{00}\left(\Omega_{x}\right)+{\mathcal{C}}_{21}\, Y_{20}
\left(\Omega_{x}\right)\,,
\label{hfic3po-rho4-text}
\end{eqnarray}
with
\begin{eqnarray}
{\mathcal{C}}_{01} & \equiv & 
-
\frac{2^{4}}{3^{5/2}}\left[1-\frac{2}{9}x^{2}\right]
e_1(x)
+\frac{2^{5}}{7^{5/2}3}
\left[1-\frac{8}{21}x^{2}\right]
 e_2(x)
\nonumber \\
{\mathcal{C}}_{21} & \equiv & 
-\, x^{2}
\left[
\frac{2^{11/2}}{3^{9/2}}\,\,
e_1(x)
-\frac{2^{17/2}}{7^{7/2}9}\,\,
e_2(x)
\right]
\label{polinomio2-text}
\end{eqnarray}
where $x=P/\beta$ and
\bea
e_1(x)=\exp\left(-\frac{x^{2}}{12}\right)
\,\,\,\,\,\,\,\,;\,\,\,\,\,\,\,\,
e_2(x)=\exp\left(-\frac{9x^{2}}{28}\right)\,.
\label{exp12}
\eea
The decay rate has a straightforward  evaluation,
by substituting (\ref{hfic3po-rho4-text}), (\ref{polinomio2-text})
in (\ref{hfi-decay}) and then in (\ref{dif-gamma})
obtaining 
\begin{equation}
\Gamma_{b_{1}\rightarrow\omega\pi}=
2\sqrt{\pi}\, x\,\frac{E_{\omega}E_{\pi}}{M_{b_{1}}}\,{\gamma}^2\,
\left({\mathcal{C}}_{01}^{2}+{\mathcal{C}}_{21}^{2}\right).
\label{taxac3-rho-5-text}
\end{equation}
The second decay process, $a_{1}\rightarrow\rho\pi$, is similar to the former
one and results in the following amplitude 
\begin{eqnarray}
{\cal M}_{fi}^{\rm C3P0} & = & {\mathcal{C}}_{01}
\, Y_{00}\left(\Omega_{x}\right)+{\mathcal{C}}_{21}
\, Y_{20}\left(\Omega_{x}\right)\,,
\label{hfic3po-rho5-text}
\end{eqnarray}
with
\begin{eqnarray}
{\mathcal{C}}_{01} & \equiv & 
\frac{2^{9/2}}{3^{5/2}}
\left[1-\frac{2}{9}x^{2}\right]
\,e_1(x)
-
\frac{2^{11/2}}{7^{5/2}\,3}
\left[1-\frac{8}{21}x^{2}\right]
\,e_2(x)
\nonumber \\
{\mathcal{C}}_{21} & \equiv & 
-\, x^{2}
\left[\frac{2^{5}}{3^{9/2}}
\,e_1(x)
-\frac{2^{7}\,5}{3^2\,7^{7/2}}
\, e_2(x)
\right]
\label{polinomio3-text}
\end{eqnarray}
and by a similar procedure one obtains
\begin{equation}
\Gamma_{a_{1}\rightarrow\rho\pi}=2\sqrt{\pi}\, 
x\,\frac{E_{\rho}E_{\pi}}{M_{a_{1}}}\,{\gamma}^2\,
\left({\mathcal{C}}_{01}^{2}
+{\mathcal{C}}_{21}^{2}\right).
\label{taxac3-rho-7-text}
\end{equation}
In the former equations, $\,e_2(x)=0$, recovers the original \3p0 results.

In addition to the decay widths $\Gamma$,  $b_1$ and $a_1$ mesons have 
$D/S$ ratios, which give a sensitive test for decay models. 
By definition, these quantities are obtained from the ratios of 
${\cal C }_{21}$ and ${\cal C }_{01} $ coefficients, in
equations (\ref{polinomio2-text}) and (\ref{polinomio3-text}).
\bea
{D\over S}\bigg|_{a_1\to\rho\pi}
 & =&
\frac{
 -\, x^{2}
\left[
\frac{2^{1/2}}{3^{2}}
e_1(x)
-\frac{3^{1/2}2^{5/2}\,5}{7^{7/2}} 
e_2(x)
\right]
}{
\left[1-\frac{2}{9}x^{2}\right]
e_1(x)
-\frac{3^{3/2}\,2 }{7^{5/2}}
\left[1-\frac{8}{21}x^{2}\right]
e_2(x)
}
\nn\\\nn\\\nn\\
{D\over S}\bigg|_{b_1\to\omega\pi}&=&
\frac{
x^{2} \left[
\frac{2^{3/2}}{3^{2}}
e_1(x)
-\frac{2^{9/2}3^{1/2}}{7^{7/2}}
e_2(x)
\right]
}{
 \left[1-\frac{2}{9}x^{2}\right]
e_1(x)
-\frac{3^{3/2}\,2}{7^{5/2}}
\left[1-\frac{8}{21}x^{2}\right]
e_2(x)
}\,.
\nn\\
\label{d/s-eq2}
\eea
The meson masses assumed in the numerical calculation were
 $M_{\pi}=138$ MeV;  
$M_{\rho}=775$ MeV;  $M_{a_1}=1230$ MeV;  $M_{b_1}=1229$ MeV;  
$M_{\omega}=782$ MeV \cite{PDG}.

The choice of SHO wave functions allow exact evaluations of the decay amplitudes even in the
corrected model. A first new aspect that appears is the presence of a new dependence in the exponential
of the corrected term. This implies in a different range for the bound-state correction due to
the fact that $e_2(x)/e_1(x)\raw0$ as $x\raw\infty$. 

The correction introduces the bound-state kernel,  Eq. (\ref{kernel}),
to the calculation of the decay processes. A general sum over the
meson index $\alpha$ is present and  as   stated before, 
this index represents  the quantum numbers of space, spin and
isospin.  The OZI-allowed decays represent, flavor conserved continuous
 (anti)quark lines. A direct consequence of this fact is the
 possibility to sum over a larger set of mesons in the $\alpha$
index. In our calculation the sum was restricted
only to the final state mesons. In the 
$b_{1}^{+}\rightarrow\omega\pi^{+}$ decay, there are two bound-state kernel
contributions one associated to $\omega$ meson and the other to $\pi^{+}$.
Similarly, the $a_{1}^{+}\rightarrow\rho^{+}\pi^{0}$ decay has two  bound-state kernel
contributions one associated to $\rho^{+}$ meson and the other to the
$\pi^{0}$. 

\begin{figure}[ht]
  \begin{center}
 \epsfig{file=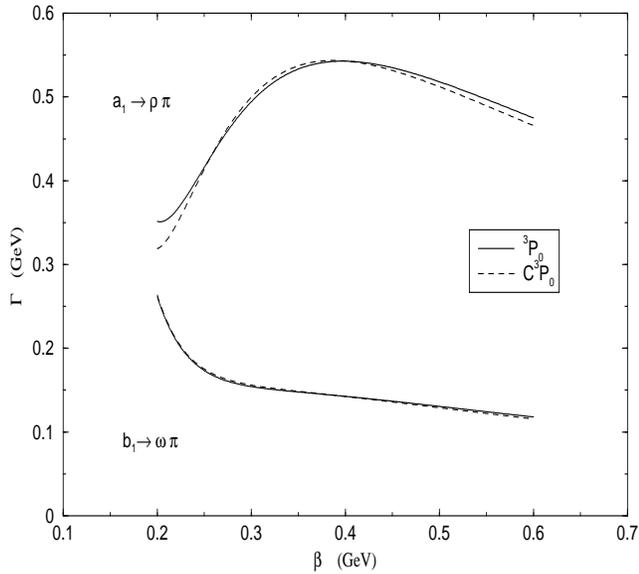, height=240.00pt, width=220.00pt,angle=-90}
  \end{center}
  \caption{Decay rates for  $b_{1}\rightarrow\omega\pi$ and
$a_{1}\rightarrow\rho\pi$ decays, for \3p0 ($\gamma=0.506$)  and C\3p0
    ($\gamma=0.539$)
    models  .  }
  \label{decays-fig}
\end{figure}

\begin{figure}[ht]
  \begin{center}
\epsfig{file=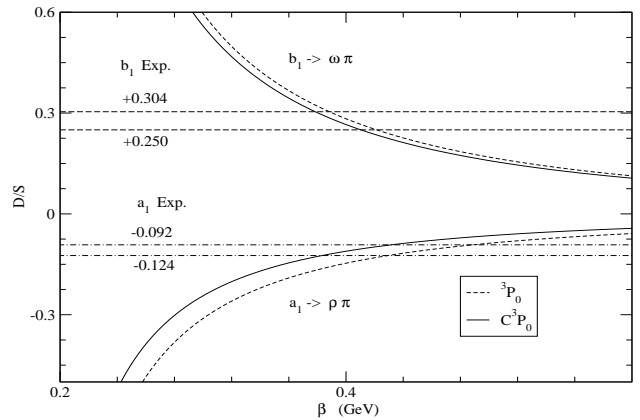, height=280.00pt, width=200.00pt, angle=-90}
  \end{center}
  \caption{$D/S$ ratios in $b_{1}\rightarrow\omega\pi$ and
$a_{1}\rightarrow\rho\pi$ decays, for \3p0 and C\3p0 models.  }
  \label{d-s}
\end{figure}

 In this example, the parameters were chosen in order to give a closer
fit to the experimental data. In the $b_1$ decay, 
width and partial waves are known with accuracy. The \3p0 model's
optimum fit for  the $b_1$ data ($\Gamma$ and $D/S$ ratio) is achieved  
with  $\gamma=0.506$ and $\beta=0.397$
GeV.  In the  C\3p0 model a similar fit is obtained with
 $\gamma=0.539$ and $\beta=0.396$ GeV.  These parameters are used in the two
models to describe the $a_1$ decay. The results for $\Gamma$ as a
function of $\beta$  appear in
figure \ref{decays-fig} and specific values are presented 
in table \ref{tab1}. In figure \ref{d-s}, the $D/S$ ratios  for the two models are 
plotted.

\begin{table*}
\begin{ruledtabular}
\caption{Decay rates
 \3p0 \rm{(}$\gamma=0.506$ e $\beta=0.397$ {\rm GeV )} and
$C^3P_0$  \rm{(}$\gamma=0.539$ e $\beta=0.396$ {\rm GeV )} }
\begin{tabular}{cc|cccc|ccccccc}
&$ $ &  &$\Gamma$ (MeV)  &   &  &  & & D/S 
&
\\
Decay &$$  &Exp \cite{PDG}  & \3p0 &C\3p0 &$$ & Exp \cite{PDG} &&\3p0 & & C\3p0 \\
  \hline
$b_1 \rightarrow \omega\pi$& & 142  &143 & 142 & &$0.277(27)$&&$0.288$ && $0.273$\\
$a_1 \rightarrow \rho\pi$& & 250 to 600&  543   & 543& & $-0.108(16)$ &&$-0.149$ & &$-0.113$ 
\label{tab1}
\end{tabular}
\end{ruledtabular}
\end{table*}

\section{Summary and Conclusions}
\label{sec:summary}

In this paper we have presented an alternative approach for meson decay
which consists  in a mapping technique, known as the Fock-Tani
formalism, long used in atomic physics. This formalism has been
applied to hadron-hadron scattering interactions with constituent
interchange. The challenge, resided in extending the approach to
include meson decay. After demonstrating that in lower order the
result obtained was equivalent to the \3p0 model, an
 additional feature pointed out  was the appearance of bound-state
corrections in the  effective decay Hamiltonian.
These corrections present a natural modification in the $q\bar{q}$
interaction strength. As an example, we studied two decay 
processes  $b_{1}\rightarrow\omega\pi$ and $a_{1}\rightarrow\rho\pi$.
 The $D/S$ ratios, in Fig. (\ref{d-s}), show that
a common range of $\beta$ values for mesons is obtained. In a new
calculation with the inclusion of other decay processes it might
require  different $\beta$ values \cite{faessler}. 
 The corrected
model presents an interesting feature that for these two mesons the
decay width differs slightly when compared with the \3p0, but  $D/S$
ratios are improved.
The examples studied here are encouraging but a 
more extensive survey
of the light meson sector would be a necessary
next step. The inclusion of the full meson octet, in the evaluation of the
 bound-state correction, may provide a fine tuning for the model.


\acknowledgements

The authors would like to thank H. St\"ocker, J. Aichelin and W. Greiner
for important and enlightening discussions. This research was supported by 
Conselho Nacional de Desenvolvimento Cient\'{\i}fico e Tecnol\'ogico (CNPq),
Universidade Federal do Rio Grande do Sul (UFRGS) and Universidade Federal de
Pelotas (UFPel).

\appendix

\section{Second and third order operators}
\label{operators}

\noindent
The second order operators
\bea
q^{(2)}_{\mu}(t)&=&\frac{1}{2}\Phi^{\ast\mu_2\nu_1}_{\alpha}
\Phi^{\mu\nu_1}_{\beta}\,M_{\alpha\beta}\,
q_{\mu_2}
\nn \\ \nn \\
\aq^{(2)}_{\nu}(t)&=&\frac{1}{2}\Phi^{\ast\mu_1\nu_2}_{\alpha}
\Phi^{\mu_1\nu}_{\beta}\,M_{\alpha\beta}\,
\aq_{\nu_2}\;,
\label{II.68}
\eea
where
\bea
M_{\alpha\beta}&=&m^{\dag}_{\alpha}M_{\beta} \sin t \cos t 
- m^{\dag}_{\alpha}m_{\beta}\sin^2t
\nn\\&&
- M^{\dag}_{\alpha}M_{\beta}(2-2\cos t -  \sin^2t) 
\nn\\
&&
- M^{\dag}_{\alpha}m_{\beta}\left(2\sin t-\sin t\cos t
\right).
\eea
The third order operators are
\bea
q^{(3)}_{\mu}(t)&=&\frac{1}{2}\Phi^{\ast\rho\sigma}_{\alpha}
\Phi^{\mu\sigma}_{\beta}\Phi^{\rho\sigma_{1}}_{\gamma}\aq^{\dag}_{\sigma_{1}}
\,M_{\alpha\beta\gamma}\,
\nn\\
&&- \frac{1}{2}\Phi^{\ast\rho\sigma}_{\alpha}
\Phi^{\mu\nu_{1}}_{\alpha}
\Phi^{\rho\sigma_{1}}_{\beta}\aq^{\dag}_{\nu_{1}}\aq^{\dag}_{\sigma_{1}}
\aq_{\sigma}\,\overline{M}_{\beta}
\nn\\
&&- \frac{1}{2}\Phi^{\ast\rho\sigma}_{\alpha}
\Phi^{\mu\nu_{1} }_{\alpha}
\Phi^{\rho_{1}\sigma}_{\beta}\aq^{\dag}_{\nu_{1}} q^{\dag}_{\rho_{1}}
q_{\rho}\overline{M}_{\beta}
\label{solq3}
\nn \\ \nn \\
\aq^{(3)}_{\nu}(t)&=&- \frac{1}{2}
\Phi^{\ast\rho\sigma}_{\alpha}\Phi^{\rho\nu}_{\beta}
\Phi^{\rho_{1}\sigma}_{\gamma}q^{\dag}_{\rho_{1}}
M_{\alpha\beta\gamma}\,\nn\\
&&+ \frac{1}{2}\Phi^{\ast\rho\sigma}_{\alpha} \Phi^{\mu_{1}
\nu}_{\alpha}
\Phi^{\rho\sigma_{1}}_{\beta}q^{\dag}_{\mu_{1}}\aq^{\dag}_{\sigma_{1}}
\aq_{\sigma}\,\overline{M}_{\beta}
\nn\\
&&+ \frac{1}{2}\Phi^{\ast\rho\sigma}_{\alpha}
\Phi^{\mu_1\nu}_{\alpha}
\Phi^{\rho_{1}\sigma}_{\beta}q^{\dag}_{\mu_{1}}
q^{\dag}_{\rho_{1}}q_{\rho}
\,\overline{M}_{\beta}\,,
\label{II.70}
\eea
where
\bea
M_{\alpha\beta\gamma}&=&
m^{\dag}_{\alpha}m_{\beta}m_{\gamma} \sin^{3}t +
M^{\dag}_{\alpha}M_{\beta}m_{\gamma}\left(\sin t-\sin^{3} t\right)
\nn\\
&&
+M^{\dag}_{\alpha}m_{\beta}M_{\gamma}\left(2\sin t-\sin t \cos t -
\sin^{3} t\right)
\nn\\
&&+
\left(M^{\dag}_{\alpha}m_{\beta}m_{\gamma} +
m^{\dag}_{\alpha}M_{\beta}m_{\gamma}\right)\left(-\cos t+\cos^{3}t\right)\nn\\
&&
+
m^{\dag}_{\alpha}m_{\beta}M_{\gamma}\left(-\cos t+\cos^{3} t +
\sin^{2} t\right)
\nn\\
&&+
M^{\dag}_{\alpha}M_{\beta}M_{\gamma}\left(2-\cos t-\cos^{3} t -\sin^{2} t
\right)\nn\\
&&+ m^{\dag}_{\alpha}M_{\beta}M_{\gamma}\left(\sin t -
\sin t \cos t - \sin^{3} t\right)
\nn\\
\overline{M}_{\beta}&=&2M_{\beta}\left(1-\cos t\right)+m_{\beta}\sin t\,.
\eea


\section{The $\delta H_m$ Hamiltonian}
\label{dhamil}

The  $\delta H_m$ Hamiltonian is evaluated from Eq. (\ref{dh}).
The $q^{\dag (3)}_{\mu}\aq^{\dag (2)}_{\nu}$
combination  can be obtained from (\ref{II.68}) and
(\ref{II.70})
\begin{eqnarray}
\!\!\!\!\!\!\!\!\!\!\!\!\!\!\!\!
\delta H_{1} &=&
 q^{\dag (3)}_{\mu}\aq^{\dag (2)}_{\nu}
\,V_{\mu\nu}
=
\delta f^{\mu\nu}_1(\alpha,\beta,\gamma)\,
V_{\mu\nu}\, m^{\dag}_{\alpha} m^{\dag}_{\beta} m_{\gamma}
\label{hoc1}
\end{eqnarray}
with
\bea
\delta f^{\mu\nu}_1(\alpha,\beta,\gamma)=
\frac{1}{4}
\Phi^{\ast\rho\sigma}_{\alpha} \Phi^{\ast\mu\tau}_{\beta}
\Delta(\rho\tau;\lambda\nu) \Phi^{\lambda\sigma}_{\gamma}.
\eea
The $q^{\dag (1)}_{\mu}\aq^{\dag (4)}_{\nu}$ combination has an
important feature: a contribution from a higher order operator. A
new generator $\tilde{M}_{\alpha}$ has to be  evaluated, with
the inclusion of the following fourth order term 
\begin{eqnarray}
\tilde{M}_{\alpha}^{\left(4\right)} & = & \frac{3}{8}
\Delta_{\alpha\beta} \Delta_{\beta\gamma} M_{\gamma} - \frac{1}{8}
M_{\beta}^{\dagger} \left[\Delta_{\alpha\gamma},
\,\Delta_{\beta\delta}\right] M_{\delta}M_{\gamma}
\nn\\
&&-
\frac{1}{4} M_{\beta}^{\dagger}
\left[M_{\alpha}, T^{\dag}_{\delta\gamma\beta }
\right]
M_{\gamma}M_{\delta}.
\end{eqnarray}
The only relevant term in the $\bar{q}_{\nu}^{\dagger(4)}$ 
for meson decay is
\bea
\bar{q}_{\nu}^{(4)}(t) &\approx &
-  \frac{1}{8} \Phi_{\alpha}^{\ast\sigma\eta}
\Delta\left(\sigma\nu;\rho\tau\right) \Phi_{\beta}^{\rho\eta}
M_{\alpha}^{\dagger\left(0\right)}(t) \bar{q}_{\tau}
M_{\beta}^{\left(0\right)}(t) .
\nn\\
\label{q_4}
\eea
The resulting contribution is then
\begin{eqnarray}
\!\!\!\!\!\!\!\!\!\!\!\!
\delta H_{2} &=&
 q^{\dag (1)}_{\mu}\aq^{\dag (4)}_{\nu}
\,V_{\mu\nu}
=
\delta f^{\mu\nu}_2(\alpha,\beta,\gamma)
V_{\mu\nu}\, m^{\dag}_{\alpha} m^{\dag}_{\beta} m_{\gamma}
\label{hoc2}
\end{eqnarray}
where
\bea
\delta f^{\mu\nu}_2(\alpha,\beta,\gamma)=
- \frac{1}{8}
\Phi^{\ast\rho\sigma}_{\alpha} \Phi^{\ast\mu\tau}_{\beta}
\Delta(\rho\tau;\lambda\nu) \Phi^{\lambda\sigma}_{\gamma}.
\eea
The $q^{\dag (5)}_{\mu}\aq^{\dag (0)}_{\nu}$ combination implies in a
fifth order generator to obtain the complete set of equations of
motion (\ref{solm&O}) and (\ref{eqq})
\begin{eqnarray}
\!\!\!\!\!\!\!\!\!\! \tilde{M}_{\alpha}^{\dagger\left(5\right)}
& =&
-M_{\beta}^{\dagger} Z_{\alpha\gamma\beta}^{\dagger} M_{\gamma}
+\frac{1}{8} M_{\omega}^{\dagger} M_{\beta}^{\dagger}
W_{\alpha\omega\beta\gamma\delta}^{\dagger} M_{\delta} M_{\gamma}
\end{eqnarray}
where
\bea
Z_{\alpha\gamma\beta}^{\dagger}
& = &
-\frac{3}{8}
T^{\dag}_{\alpha\delta\beta}
\Delta_{\delta\gamma}
-\frac{5}{8}
\Delta_{\beta\delta}
T^{\dag}_{\alpha\gamma\delta}
-\frac{1}{4}
T^{\dag}_{\delta\gamma\beta}
\Delta_{\delta\alpha}
\nn\\
W_{\alpha\omega\beta\gamma\delta}^{\dagger}
& = &
\left[M_{\alpha}^{\dagger},\,
Q_{\omega\beta\gamma\delta}\right]
-\left[
\Delta_{\omega\gamma},\,
T^{\dag}_{\alpha\delta\beta}
\right]
\nn\\
Q_{\alpha\beta\gamma\delta}
&=& -\frac{1}{2}
\left[\Delta_{\alpha\gamma}, \,
\Delta_{\beta\delta}\right]
-
\left[M_{\alpha},
T^{\dag}_{\delta\gamma\beta}
\right].
\eea
The only relevant terms in the $q_{\mu}^{\dagger(5)}$ for meson decay are 
\bea
q_{\mu}^{\dagger(5)}(t) & \approx & 
\left[\frac{1}{2}
\Delta\left(\rho\tau;\mu\omega\right)
\Phi_{\alpha}^{\ast\rho\sigma} \Phi_{\beta}^{\ast\lambda\tau}
\Phi_{\gamma}^{\lambda\sigma} 
\right.
\nn\\
&& -  \frac{1}{4} \Delta\left(\rho\tau;\mu\sigma\right)
\Phi_{\alpha}^{\ast\rho\omega} \Phi_{\beta}^{\ast\lambda\tau}
\Phi_{\gamma}^{\lambda\sigma}
\nn\\
&&
\left.
 -  \frac{3}{8} \Delta\left(\rho\tau;\lambda\omega\right)
\Phi_{\alpha}^{\ast\mu\tau} \Phi_{\beta}^{\ast\rho\sigma}
\Phi_{\gamma}^{\lambda\sigma}
\right]
\nn\\
&&
\times 
 M_{\alpha}^{\dagger\left(0\right)}(t)
M_{\beta}^{\dagger\left(0\right)}(t) \bar{q}_{\omega}
M_{\gamma}^{\left(0\right)}(t).
\label{q_5}
\eea
The resulting contribution is
\bea
\delta H_3&=&
\delta f^{\mu\nu}_3(\alpha,\beta,\gamma)
V_{\mu\nu}\,\,m_{\alpha}^{\dagger}m_{\beta}^{\dagger}m_{\gamma}
\label{hco_3}
\eea
where
\bea
\delta f^{\mu\nu}_3(\alpha,\beta,\gamma)&=&
\frac{1}{2}\Phi_{\alpha}^{\ast\rho\sigma}
\Phi_{\beta}^{\ast\lambda\tau}
\Delta(\rho\tau;\mu\nu)
\Phi_{\gamma}^{\lambda\sigma}
\nn\\
&&
-
\frac{1}{4}
\Phi_{\alpha}^{\ast\sigma\tau}
\Phi_{\beta}^{\ast\rho\nu}
\Delta(\rho\tau;\mu\lambda)
\Phi_{\gamma}^{\sigma\lambda}
\nn\\
&&
- \frac{3}{8}
\Phi_{\alpha}^{\ast\rho\sigma}
\Phi_{\beta}^{\ast\mu\tau}
\Delta(\rho\tau;\lambda\nu)
\Phi_{\gamma}^{\lambda\sigma}.
\eea
%
%
%
The complete $\delta H_m$ Hamiltonian is
\bea
\delta H_m&=&\delta H_1 + \delta H_2 +  \delta H_3
\nn\\
&=&
 \delta f^{\mu\nu}(\alpha,\beta,\gamma)
V_{\mu\nu}\,\,m_{\alpha}^{\dagger}m_{\beta}^{\dagger}m_{\gamma}
\eea
with
\bea
\!\!\!\!\!\!\!\!\!\!\!\!\!\!\!\!\!\!\!\!\!\!\!\!
\delta f^{\mu\nu}(\alpha,\beta,\gamma)&=&
\delta f^{\mu\nu}_1
+\delta f^{\mu\nu}_2
+\delta f^{\mu\nu}_3
\nn\\
&=&
\,\,\,\,\,
\frac{1}{2}\Phi_{\alpha}^{\ast\rho\sigma}
\Phi_{\beta}^{\ast\lambda\tau}
\Delta(\rho\tau;\mu\nu)
\Phi_{\gamma}^{\lambda\sigma}
\nn\\
&&
-
\frac{1}{4}
\Phi_{\alpha}^{\ast\rho\sigma}
\Phi_{\beta}^{\ast\mu\tau}
\Delta(\rho\tau;\lambda\nu)
\Phi_{\gamma}^{\lambda\sigma}.
\nn\\
&&
-
\frac{1}{4}
\Phi_{\alpha}^{\ast\sigma\tau}
\Phi_{\beta}^{\ast\rho\nu}
\Delta(\rho\tau;\mu\lambda)
\Phi_{\gamma}^{\sigma\lambda}.
\label{hco_fim}
\eea

\section{  Decay, Wave function and matrix elements}
\label{wave-matrix}

 We will use 
the decay $b_1^{+}(+\hat{z})\rightarrow\omega(+\hat{z})\pi^{+}$ to illustrate
the nature of our formalism and, simply quote the other case in the
text.

\subsection{The wave function}

\noindent
The general meson wave function can be written as 
\bea
\Phi_{\alpha}^{\mu\nu}=
\chi_{S_\alpha }^{ s_{1}s_{2} } 
f_{f_{\alpha} }^{f_{1}f_{2}}
C^{c_{1}c_{2}}
\Phi_{nl }^{\vec{P}_{\alpha}-\vec{p}_{1}-\vec{p}_{2}}
\;,
\label{funcdomeson}
\eea
a   direct
product of the spin  $\chi_{S_{\alpha}}^{s_{1}s_{2}}$
[the indexes $s_{1}$ and $s_{2}$ are
the quark (antiquark) spin projections
with $(s=1 \Rightarrow \uparrow $ and
$s=2\Rightarrow\downarrow)$; the index
$S_{\alpha} $ denotes the meson spin]; 
flavor $f_{f_{\alpha}}^{f_{1}f_{2}}$; color $C^{c_{1}c_{2}}$ and
space $\Phi_{nl }^{\vec{P}_{\alpha}-\vec{p}_{1}-\vec{p}_{2}}$ components.

In all our calculations the color component will be given by
\bea
C^{c_{1}c_{2}}=\frac{1}{\sqrt{3}}\,\,\delta^{c_{1}c_{2}}.
\label{color}
\eea
The spatial part is defined as harmonic oscillator wave functions
\bea
\Phi_{nl }^{\vec{P}_{\alpha}-\vec{p}_{1}-\vec{p}_{2}}
=
\delta(\vec{P}_{\alpha}-\vec{p}_{1}-\vec{p}_{2})
\,\,\Phi_{nl}(\vec{p}_1,\vec{p}_2)
\eea
where
$\Phi_{nl}(\vec{p}_i,\vec{p}_j)  $ is given by
\bea
\Phi_{nl}(\vec{p}_i,\vec{p}_j)&=&
 (\frac{1}{2\beta})^{l}\, N_{nl}\, |\vec{p}_i-\vec{p}_j|^{l}\,
\exp\left[ -\frac{(\vec{p}_i-\vec{p}_j)^2   }{ 8\beta^2 }   \right]\,
\nn\\
&&\times {\cal L}_{n}^{l+\frac{1}{2}}
\left[\frac{(\vec{p}_i-\vec{p}_j)^2   }{ 4\beta^2} \right] 
 Y_{lm}(\Omega_{\vec{p}_i-\vec{p}_j}     ) 
\label{psi_oh}
\eea
with $p_{i(j)} $ the internal momentum, the spherical harmonic $Y_{lm}$, 
$\beta$ a scale parameter, $N_{nl}$ the normalization constant dependent
on the radial and orbital quantum numbers
\bea
N_{nl}=
\left[   
\frac{2 (n!)}{\beta^3\,\Gamma(n+l+3/2) }
\right]^{\frac{1}{2}}.
\eea
The Laguerre polynomials ${\cal L}_{n}^{l+\frac{1}{2}}(p)$ are defined as
\bea
{\cal L}_{n}^{l+\frac{1}{2}}(p)=\sum_{k=0}^{n}
\frac{(-)^k\, \Gamma(n+l+3/2)^{(n-k)!}  }{ k!\, \Gamma(k+l+3/2)  }\,\,p^k\,.
\label{laguerre}
\eea
In this paper two kinds of light non-strange mesons will be studied:
\begin{enumerate}
 \item  $L_{q\bar{q}}=0$
\bea
\vphi(\vec{p})\equiv \Phi_{00}(\vec{p})=\frac{1}{\pi^{3/4}\beta^{3/2}} 
\exp\left[ -\frac{p^2   }{ 8\beta^2 }   \right]\,
\label{phi-L0}
\eea
 \item $L_{q\bar{q}}=1$
\bea
\Phi_{1m}(\vec{p})=
\phi(\vec{p})\,\, Y_{1m} (\Omega_{\vec{p}})
\label{phi-L1}
\eea
\end{enumerate}
with
\bea
 \phi(\vec{p})= \left[\frac{2}{3\sqrt{\pi}\beta^{5} } \right]^{\frac{1}{2}}
\,\, p \, \,
\exp\left[ -\frac{p^2   }{ 8\beta^2 }   \right].
\eea
Returning to our example the pion, has
 $J=0$ and $b_1$  $J=1$. We choose the  $\left(+\hat{z}\right)$
direction for this calculation, so the spin wave functions become
\bea
\left|b_1\right\rangle  & = & \frac{1}{\sqrt{2}}
\left(\left|\uparrow\bar{\downarrow}\right\rangle -
\left|\downarrow\bar{\uparrow}\right\rangle \right)
\nn\\
\left|\omega\right\rangle  & = &
\left|\uparrow\bar{\uparrow}\,\right\rangle 
\nn\\
\left|\pi\right\rangle  & = & \frac{1}{\sqrt{2}}
\left(\left|\uparrow\bar{\downarrow}\right\rangle -
\left|\downarrow\bar{\uparrow}\right\rangle \right)
\eea
or in the $\chi$ notation
\bea
&&\!\!\!\!\!\!\!\!\!\!\!\!\!\!\!\!
\chi_{\omega}^{11}  =  1,
\quad\,\chi_{\omega}^{12}= \chi_{\omega}^{21}= \chi_{\omega}^{22}=0 
\nn \\
&&\!\!\!\!\!\!\!\!\!\!\!\!\!\!\!\!
\chi_{\pi,b_1}^{11} =    \chi_{\pi,b_1}^{22}= 0
\,\,\,\,\,\,;\,\,\,\,\,\,
\chi_{\pi,b_1}^{12}=- \chi_{\pi,b_1}^{21}=   \frac{1}{\sqrt{2}}.
\eea
The flavor component $f_{f_{\alpha}}^{f_{\mu}f_{\nu}}$  follows the
same logic as the spin part
\begin{eqnarray}
\left|b_1^{\,+}\,\right\rangle & = &
\left|\pi^{+}\right\rangle=
 -\left|u\bar{d}\right\rangle
 \nonumber \\
\left|\omega\,\right\rangle  & = &
\frac{1}{\sqrt{2}}\left(\left|u\bar{u}\right\rangle
+\left|d\bar{d}\right\rangle \right)
\end{eqnarray}
The  $b_1^+$ and the $\pi^+$ mesons have the same flavor contribution
\begin{eqnarray}
\!\!\!\!\!\!\!\!\!\!
f_{b_1^{+},\pi^{+}}^{12}=-1\,;
f_{b_1^{+},\pi^{+}   }^{11} & =
f_{b_1^{+},\pi^{+}  }^{21}=f_{b_1^{+},\pi^{+} }^{22}= & 0\,.
\end{eqnarray}
For $\omega$, one has
\begin{eqnarray}
f_{\omega}^{11} & =f_{\omega}^{22}= & \frac{1}{\sqrt{2}};
\qquad f_{\omega}^{12}=f_{\omega}^{21}=0.
\end{eqnarray}

\subsection{The spin matrix elements}

In the evaluation of a decay amplitude, the following spin matrix element is necessary
\begin{equation}
\chi_{s^{\prime}}^{\ast}
\left(\vec{\sigma}\cdot\vec{ {P}}\right)\chi_{s}^{c}
\label{e4}
\end{equation}
with
\begin{equation}
\chi_{1}=\left(\begin{array}{c}
1\\
0\end{array}\right);\;\chi_{2}=\left(\begin{array}{c}
0\\
1\end{array}\right);\;\chi_{1}^{c}=\left(\begin{array}{c}
0\\
1\end{array}\right);\;\chi_{2}^{c}=\left(\begin{array}{c}
-1\\
0\end{array}\right)
\label{e2}.
\end{equation}
By direct calculation one can show
\begin{eqnarray}
\chi_{1}^{\ast}\left(\vec{\sigma}\cdot\vec{ {P}}\right)\chi_{1}^{c} & = &  {P}_{x}-i {P}_{y}
\nn\\
\chi_{1}^{\ast}\left(\vec{\sigma}\cdot\vec{ {P}}\right)\chi_{2}^{c} & = &  -{P}_{z}
\nn\\
\chi_{2}^{\ast}\left(\vec{\sigma}\cdot\vec{ {P}}\right)\chi_{1}^{c} & = &  -{P}_{z}
\nn\\
\chi_{2}^{\ast}\left(\vec{\sigma}\cdot\vec{ {P}}\right)\chi_{2}^{c} & = & - ({P}_{x}+i {P}_{y})\,.
\label{e10}
\end{eqnarray}

\subsection{Matrix elements: $b_{1}^{+}\rightarrow\omega\pi^{+}$ Decay  }

The transition considered is of the form
$m_{\gamma}\rightarrow m_{\alpha}+m_{\beta}$, where the 
initial state is
$\left|A \right\rangle   =  m_{\gamma}^{\dagger}
\left|0\right\rangle $ and the
 final state is given by  $
\left|B C\right\rangle   =
m_{\alpha}^{\dagger}m_{\beta}^{\dagger}
\left|0\right\rangle $.
The matrix element of the uncorrected part 
results in
\begin{eqnarray}
\left\langle B C \left|H_{m}\right|A\right\rangle =-d_{1}-d_{2},
\label{4.1-4a}
\end{eqnarray}
$d_{1}$ and $d_{2}$ are defined as
\begin{eqnarray}
d_{1}&\equiv&\Phi_{\alpha}^{\star\rho\nu}\Phi_{\beta}^{\star\mu\lambda}
\Phi_{\gamma}^{\rho\lambda}V_{\mu\nu}
\nonumber \\
d_{2}&\equiv&   \Phi_{\alpha}^{\star\mu\lambda}
\Phi_{\beta}^{\star\rho\nu}\Phi_{\gamma}^{\rho\lambda}
V_{\mu\nu}.
\label{4.1-4b}
\end{eqnarray}
Equations (\ref{4.1-4b}) can be decomposed according to the sector of
the wave function they correspond: flavor, color, spin-space:
\begin{eqnarray}
d_{1}&=& d_{1}^{f}\,d_{1}^{c}\,d_{1}^{s-e} \nonumber \\
d_{2}&=& d_{2}^{f}\,d_{2}^{c}\,d_{2}^{s-e}.
\end{eqnarray}

The matrix elements of the bound-state correction refer to diagrams
(b), (c) and (d) of figure (\ref{diag}). 
The bound-state kernel's definition 
as $\Phi^{\rho\tau}_{\alpha}\Phi^{\ast\lambda\nu}_{\alpha}$
implies in an additional element, due to the contraction in the
$\alpha$ index, a {\it sum over species}
requirement \cite{girar1}.  A question that naturally arises is,  which
states to include in this sum?  We shall adopt in our 
calculation a restrictive choice: include in the sum only
the particles that are present in the final state. 
For the $b_1^{+}$ decay,  $\Delta\left(\rho\tau;\lambda\nu\right)$
will have two contributions: $\omega$ and $\pi^{+}$. Similarly, 
the $a_1^{+}$ decay shall be corrected by the final state mesons $\rho^{+}$ and $\pi^{0}$. 

Due to the parity assignment
of the spatial part, the integration of diagram  (\ref{diag}b) is zero. Spatial
symmetry also implies that the matrix elements of diagrams  (\ref{diag}c) and (\ref{diag}d)
are equal. This simplifies our calculation, reducing the problem to
the evaluation of diagram  (\ref{diag}c) only. 
 The bound-state correction (bsc) matrix element reduces to evaluate
 the following expression
\begin{eqnarray}
\left\langle B C\left|\delta H_m\right|A\right\rangle
=-d_{1}^{\rm bsc}-d_{2}^{\rm bsc}
\label{4.3-4}\end{eqnarray}
where 
\begin{eqnarray}
d_{1}^{\rm bsc} & = & \frac{1}{4}\left(d_{1\omega} +d_{1\pi}\right)\nonumber \\
d_{2}^{\rm bsc} & = & \frac{1}{4}\left(d_{2\omega} +d_{2\pi}\right)
\label{4.3-5c}
\end{eqnarray}
with
\begin{eqnarray}
d_{1j } & = & \Phi_{\alpha}^{\star\rho\sigma}
\Phi_{\beta}^{\star\mu\tau} \Delta_{j}(\rho\tau;\lambda\nu)
\Phi_{\gamma}^{\lambda\sigma}V_{\mu\nu}\;
\equiv\;d_{1j}^{f}d_{1j}^{c} d_{1j}^{s-e}\nonumber \\
d_{2j} & = & \Phi_{\alpha}^{\star\mu\tau}
\Phi_{\beta}^{\star\rho\sigma} \Delta_{j}(\rho\tau;\lambda\nu)
\Phi_{\gamma}^{\lambda\sigma}V_{\mu\nu}\;
\equiv\;d_{2j}^{f}d_{2j}^{c} d_{2j}^{s-e}.
\nonumber \\
\label{4.3-5d}
\end{eqnarray}
In (\ref{4.3-5d}) $j$ refers to mesons $\omega$ and $\pi^{+}$.

\subsection{$b_{1}^{+}\rightarrow\omega\pi^{+}$ Decay (uncorrected)}

$\bullet$ Flavor:
\label{sub:sabor-b1}

\begin{eqnarray}
d_{1}^{\, f} & = &
d_{2}^{\, f}=
f_{\pi}^{\, f_{\rho}f_{\nu}}f_{\omega}^{\!f_{\mu}f_{\lambda}}
f_{b_1 }^{\! f_{\rho}f_{\lambda}}\delta_{f_{\mu}f_{\nu}}
=\frac{1}{\sqrt{2}}.
\label{4.2-6}
\end{eqnarray}

$\bullet$ Color:

\begin{eqnarray}
\!\!\!\!\!\!
d_{1}^{c} = 
d_{2}^{c}=
\frac{1}{3\sqrt{3}}\delta^{\, c_{\rho}c_{\nu}}
\delta^{ c_{\mu}c_{\lambda}}
\delta^{ c_{\rho}c_{\lambda}}\delta^{c_{\mu}c_{\nu}}
=\frac{1}{\sqrt{3}}.
\label{4.1-25}
\end{eqnarray}

$\bullet$ Spin-space:

The spin matrix element is
\begin{eqnarray}
d_{1}^{\, s} & = & 
d_{2}^{\, s}=
\chi_{\pi}^{\, s_{\rho}s_{\nu}}\chi_{\omega}^{\!
s_{\mu}s_{\lambda}}
\chi_{b_1}^{\!
s_{\rho}s_{\lambda}}V_{s_{\mu}s_{\nu}}^{s-e}
=\frac{1}{2}\,V_{11}^{s-e}\left(\vec{p_\mu},\vec{p_\nu}\right)
\nn\\
\label{4.2-12}
\end{eqnarray}
where
\bea
\!\!\!\!\!\!\!\!\!\!\!\!
V_{11}^{s-e}\left(\vec{p_\mu},\vec{p_\nu}\right)=-\gamma\,
\delta(\vec{p_\mu}+\vec{p_\nu})
\chi^{\ast}_{1}\,[\vec{\sigma}\cdot (\vec{p_\mu}-\vec{p_\nu})]\,
\chi^{C}_{1}.
\label{v_spin}
\eea
Using (\ref{e10}) to evaluate (\ref{v_spin}) and
after integrating   momentum conservation deltas one arrives in
\begin{eqnarray}
d_{1}^{s-e} & = & - \gamma\,\,
\int d^{3} {K}\:\left( {K}_{x}-i
{K}_{y}\right)
\varphi\left( {\vec{P}}-2\vec{ {K}}\right)
\nonumber \\
 & \times &
\phi\left(2 {\vec{P}}-2\vec{ {K}}\right)Y_{11}
\left(\Omega_{2 {\vec{P}}\hat{-}2\vec{ {K}}} \right)\,
\varphi\left(
{\vec{P}}-2\vec{ {K}}\right)\,.
\nn\\
\label{4.2-14}
\end{eqnarray}
Introducing the spatial wave function and integrating 
\begin{eqnarray}
d_{1}^{\, s-e} & = & \left(\frac{2^{7/2}}{3^{5/2}}\right)
\,\left(\frac{\gamma}{\pi^{1/4}\,\beta^{1/2}}\right)
  \left\{ \left[1-\frac{2}{9}
  x^2\right]Y_{00}\left(\Omega_x\right)
\right.
\nn\\
&&
\left.
+\frac{2}{3^{2}\sqrt{5}}
x^2\,
Y_{20}\left(\Omega_x\right)\right\} e_1(x).
\label{4.2-24}
\end{eqnarray}
 $d_{2}^{\, s-e}$ is obtained from  $d_{1}^{\, s-e}$ by
 $\vec{P}\to-\vec{P}$. The decay amplitude results
\begin{eqnarray}
h_{fi} & = & -\left(\frac{2^{4}}{3^{5/2}}\right)\,
\left(\frac{\gamma}{\pi^{1/4}\,\beta^{1/2}}\right)
\left\{ \left[1-\frac{2}{9}x^2
\right]Y_{00}\left(\Omega_x\right)
\right.
\nn\\
&&
\left.
+\frac{2}{3^{2}\sqrt{5}}x^2\,
Y_{20}\left(\Omega_x\right)\right\} e_1(x).
\label{4.2-27}
\end{eqnarray}

\subsection{$b_{1}^{+}\rightarrow\omega\pi^{+}$ Decay (bound-state corrected)}

The quantities between $[\ldots]$ in the following expressions are
related to the bound-state kernel.

$\bullet$ Flavor:

\begin{eqnarray}
d_{1\omega}^{\, f} & = & f_{\pi}^{\, f_{\rho}f_{\sigma}}f_{\omega}^{\!
  f_{\mu}f_{\tau}}
\left[f_{\omega}^{\, f_{\rho}f_{\tau}}f_{\omega}^{\!f_{\lambda}f_{\mu}}\right]
f_{b_1 }^{\! f_{\lambda}f_{\sigma}}
=\frac{1}{2\sqrt{2}}
\nonumber \\
d_{1\pi}^{\, f} & = & f_{\pi}^{\, f_{\rho}f_{\sigma}}f_{\omega}^{\!
  f_{\mu}f_{\tau}}
\left[f_{\pi}^{\, f_{\rho}f_{\tau}}f_{\pi}^{\!f_{\lambda}f_{\mu}}\right]
f_{b_1 }^{\! f_{\lambda}f_{\sigma}}
=\frac{1}{\sqrt{2}}\nonumber \\
d_{2\omega}^{\, f} & = & f_{\pi}^{\! f_{\mu}f_{\tau}}f_{\omega}^{\,f_{\rho}f_{\sigma}}
\left[f_{\omega}^{\, f_{\rho}f_{\tau}}f_{\omega}^{\!f_{\lambda}f_{\mu}}\right]
f_{b_1 }^{\! f_{\lambda}f_{\sigma}}
=\frac{1}{2\sqrt{2}}\nonumber \\
d_{2\pi}^{\, f} & = & f_{\pi}^{\!
  f_{\mu}f_{\tau}}f_{\omega}^{\,f_{\rho}f_{\sigma}}
\left[f_{\pi}^{\, f_{\rho}f_{\tau}}f_{\pi}^{\!f_{\lambda}f_{\mu}}\right]
f_{b_1 }^{\!f_{\lambda}f_{\sigma}}=0.
\label{4.3-70}
\end{eqnarray}

$\bullet$ Color:

\begin{eqnarray}
d_{1\omega}^{\, c} & = & d_{1\pi}^{\, c}=d_{2\omega}^{\, c}=d_{2\pi}^{\, c}
\nn\\
&=&
\frac{1}{9\sqrt{3}}
\delta^{\,c_{\rho}c_{\sigma}}
\delta^{c_{\mu}c_{\tau}}
[\delta^{ c_{\rho}c_{\tau}}
\delta^{ c_{\lambda}c_{\nu}}]
\delta^{ c_{\lambda}c_{\sigma}}
\delta^{c_{\mu}c_{\nu}}
\nn\\
 &=& \frac{1}{3\sqrt{3}}.
\label{4.3-8c}
\end{eqnarray}

$\bullet$ Spin-space:

The spin matrix element is
\begin{eqnarray}
d_{1\omega}^{\, s} & =& \chi_{\pi}^{\,
  s_{\rho}s_{\sigma}}\chi_{\omega}^{\! s_{\mu}s_{\tau}}
\left[\chi_{\omega}^{\, s_{\rho}s_{\tau}}\chi_{\omega}^{\! s_{\lambda}s_{\nu}}\right]
\chi_{b_1 }^{\! s_{\lambda}s_{\sigma}}
V_{s_{\mu}s_{\nu}}^{s-e}
\nn\\
&=&\frac{1}{2}V_{11}^{s-e}\left(\vec{p_\mu},\vec{p_\nu}\right)
\nonumber \\
d_{1\pi}^{\, s} &= & 
\chi_{\pi}^{\, s_{\rho}s_{\sigma}}\chi_{\omega}^{\! s_{\mu}s_{\tau}}
\left[\chi_{\pi}^{\, s_{\rho}s_{\tau}}\chi_{\pi}^{\! s_{\lambda}s_{\nu}}\right]
\chi_{b_1 }^{\! s_{\lambda}s_{\sigma}}
V_{s_{\mu}s_{\nu}}^{s-e}
\nn\\
&=&\frac{1}{4}V_{11}^{s-e}\left(\vec{p_\mu},\vec{p_\nu}\right)
\nonumber \\
d_{2\omega}^{\, s} & = &
\chi_{\pi}^{\! s_{\mu}s_{\tau}}\chi_{\omega}^{\, s_{\rho}s_{\sigma}}
\left[\chi_{\omega}^{\, s_{\rho}s_{\tau}}\chi_{\omega}^{\! s_{\lambda}s_{\nu}}\right]
\chi_{b_1 }^{\! s_{\lambda}s_{\sigma}}V_{s_{\mu}s_{\nu}}^{s-e}
\nn\\
&=& 0
\nonumber \\
d_{2\pi}^{\, s} & = &
\chi_{\pi}^{\! s_{\mu}s_{\tau}}\chi_{\omega}^{\,s_{\rho}s_{\sigma}}
\left[\chi_{\pi}^{\, s_{\rho}s_{\tau}}\chi_{\pi}^{\!s_{\lambda}s_{\nu}}\right]
\chi_{b_1 }^{\!s_{\lambda}s_{\sigma}}
V_{s_{\mu}s_{\nu}}^{s-e}
\nn\\
&=&\frac{1}{4}V_{11}^{s-e}\left(\vec{p_\mu},\vec{p_\nu}\right).
\label{4.3-10}
\end{eqnarray}
Due to symmetries in the spatial part the following relations are true
\begin{eqnarray}
 d_{1\pi}^{\, s-e}&=&d_{2\pi}^{\, s-e}
=\frac{1}{2} \,d_{1\omega}^{\, s-e};
\qquad d_{2\omega}^{\, s-e}=0\,,
\label{4.3-72}
\end{eqnarray}
where  $d_{1\pi}^{\, s-e}$ is given by
\begin{eqnarray}
d_{1\pi}^{s-e} & = & \frac{\gamma}{2}
\int d^{3}K\,d^{3}q 
\,(K_x-i\,K_y)
\nonumber \\
& &\!\!\! \times  \varphi\left(2\vec{q}-\vec{ {P}}\right)
\varphi\left(2\vec{K}-\vec{ {P}}\right)
\nonumber \\
& &\!\!\! \times 
\left[\varphi\left(\vec{q} +\vec{K} -2\vec{{P}}\right)
\varphi\left(\vec{q} +\vec{K}\right)\right]
\phi\left(2\vec{q}\right)Y_{11}\left(\Omega_{2\vec{q}}\right).
\nn\\
\label{4.3-75}
\end{eqnarray}
After integration one finds
\begin{eqnarray}
d_{1\pi}^{\, s-e} & = &\!\!\!-
\left(\frac{2^{11/2}}{7^{5/2}}\right)
\,\left(\frac{\gamma}{\pi^{1/4}\,\beta^{1/2}}\right)
 \left\{\left[1-\frac{8}{21}x^2 \right]Y_{00}\left(\Omega_x\right)
\right.
\nn\\
&&\left.
+\frac{2^{7/2}}{21}\sqrt{\frac{1}{10}}\,x^2
Y_{20}\left(\Omega_x\right)\right\}e_2(x) .
\label{4.3-89}
\end{eqnarray}
The decay amplitude for the bound-state correction
\begin{eqnarray}
h_{fi}^{\rm bsc} & = & \left(\frac{2^{4}}{7^{5/2}3}\right)\,
\left(\frac{\gamma}{\pi^{1/4}\,\beta^{1/2}}\right)\,
\left\{\left[1-\frac{8}{21}\,x^2\right]Y_{00}
\left(\Omega_x\right)
\right.
\nn\\
&&
\left.
+\frac{2^{7/2}}{21}\, x^{2}Y_{20}
\left(\Omega_x\right)\right\}\,e_2(x).
\label{4.3-95}
\end{eqnarray}
The total amplitude will be
\bea
h_{fi}^{\rm C3P0}=h_{fi}+2\,h_{fi}^{\rm bsc}= \frac{\gamma}{\pi^{1/4}\,\beta^{1/2}}
\,{\cal  M}_{fi}^{\rm C3P0},
\label{ampli-tot}
\eea
which are  expressions (\ref{hfi-decay}) and
(\ref{hfic3po-rho4-text}).

\end{document}